\documentclass[a4paper, 12pt]{article}

\usepackage[sort&compress]{natbib}
\bibpunct{(}{)}{;}{a}{}{,} 

\usepackage{amsthm, amsmath, amssymb, mathrsfs, multirow, url, subfigure}
\usepackage{array}
\usepackage{graphicx} 
\usepackage{ifthen} 
\usepackage{amsfonts}
\usepackage[usenames]{color}
\usepackage{fullpage}
\usepackage{algorithm, algorithmicx, algpseudocode}
\usepackage{bm}
\usepackage{multirow}

\theoremstyle{plain}

\theoremstyle{definition}

\theoremstyle{remark}

\newcommand{\nm}{{\sf N}}

\newcommand{\chisq}{{\sf ChiSq}}

\newcommand{\RR}{\mathbb{R}}

\newcommand{\PP}{\mathbb{P}}

\newcommand{\veta}{\boldsymbol{\veta}}

\renewcommand{\phi}{\varphi} 
\newcommand{\eps}{\varepsilon}

\newcommand{\iid}{\overset{\text{\tiny iid}}{\,\sim\,}}

\title{Calibrating general posterior credible regions}
\author{
Nicholas Syring\footnote{Department of Mathematics, Statistics, and Computer Science, University of Illinois at Chicago} \quad and \quad Ryan Martin\footnote{Department of Statistics, North Carolina State University, {\tt rgmarti3@ncsu.edu}}
}
\date{\today}

\begin{document}

\maketitle 

\begin{abstract}
An advantage of methods that base inference on a posterior distribution is that credible regions are readily obtained.  Except in well-specified situations, however, there is no guarantee that such regions will achieve the nominal frequentist coverage probability, even approximately.  To overcome this difficulty, we propose a general strategy that introduces an additional scalar tuning parameter to control the posterior spread, and we develop an algorithm that chooses this parameter so that the corresponding credible region achieves the nominal coverage probability.  %Simulations demonstrate effectiveness of the proposed method.  
	
\smallskip

\end{abstract}

\section{Introduction}
\label{S:intro}

An advantage of methods that base their inference on a posterior distribution is that credible regions for the unknown parameters are readily available.  It is common to require that the specified credibility level agrees, at least approximately, with the frequentist coverage probability, e.g., that the 95\% credibility regions from the posterior are approximately 95\% confidence regions.  If so, then we say that the posterior credible region is \emph{calibrated}.  For well-specified Bayesian models, a Bernstein--von Mises theorem justifies a calibration claim, but when the model is misspecified, calibration often fails.  For example, \citet{kleijn.2012} derived a Bernstein--von Mises theorem for Bayesian posteriors under model misspecification, and pointed out that, even if the concentration target and rate are correct, misspecification can still cause a lack of calibration; see page~362 in their paper.  Similarly, variational Bayes posteriors \citep[e.g.,][]{blei.VI} often lack the calibration property, and correcting this is an important open problem.  

To address this problem, we propose to augment the given posterior with an additional scalar tuning parameter that controls its spread.  This is inspired by the literature on Gibbs posteriors, where data and the parameter of interest are connected via a loss function, like in \eqref{eq:Gibbs}, instead of a likelihood; see, e.g., \citet{bissiri.holmes.walker.2013}, \citet{alquier.2015}, \citet{zhang2006}, \citet{jiang.tanner.2008}, and \citet{syring.martin.mcid}.  In such cases, a scale parameter must be specified to properly weight the information in the data relative to that in the prior, but this boils down to tuning the spread.  A similar formulation is possible for other types of posterior distributions, not just Gibbs; see Section~\ref{S:problem}.  Having introduced an extra parameter, we propose to select its value so that the corresponding posterior credible regions are calibrated in the sense described above, and we present a Monte Carlo algorithm to implement this idea.  

Similar questions about scaling posterior distributions to address model misspecification have been considered recently in the literature.  Ideas for choosing the posterior scaling are presented in \citet{bissiri.holmes.walker.2013} and \citet{holmes.walker.2016}, including hierarchical Bayes and loss/information matching, and a novel idea in \citet{grunwald.ommen.scaling}.  These proposals are reasonable, but they provide no guarantee that the corresponding posterior uncertainty quantification is meaningful.  In contrast, our proposal here is designed specifically to calibrate  posterior credible regions, at least approximately.  Theoretical arguments support the soundness of our proposal, and numerical examples demonstrate both its versatility and effectiveness.  Finer details about implementation of the method, along with two additional examples, are presented in the Supplementary Material.

%The remainder of the paper is organized as follows.  Section~\ref{S:problem} sets our notation, defines our modified posterior distribution, with an extra {calibration parameter}, and explains the intuition behind our proposed approach.  The \emph{general posterior calibration} algorithm is presented in Section~\ref{S:algorithm} and we discuss its basic properties.  Section~\ref{S:examples} contains several examples, including a Gibbs posterior in quantile regression, a misspecified Bayes posterior in linear regression, and a variational Bayes posterior in a mixture model, and Section~\ref{S:discuss} makes some concluding remarks.                 

\section{Problem formulation}
\label{S:problem}

Suppose we have data $Z^n=(Z_1,\ldots,Z_n)$ consisting of independent and identically distributed observations with marginal distribution $P$; each $Z_i$ could be a vector or even a response--predictor variable pair, i.e., $Z_i = (X_i, Y_i)$.  Throughout, we write $Pf$ for the expectation of a function $f(Z)$ with respect to $P$.  The quantity of interest is $\theta=\theta(P)$, a feature of $P$, taking values in $\Theta \subseteq \RR^d$ for some $d \geq 1$.  

Consider the following general construction of a posterior distribution for inference on $\theta$.  Start by connecting data $Z^n$ to a full set of parameters $\eta$ either through a statistical model for $P$, as in Bayesian settings, or through a suitable loss function, as in Gibbsian settings like in \eqref{eq:Gibbs}. Next, introduce a prior $\Pi$ for the full parameter $\eta$, and a scale $\omega > 0$ to weight the information about $\eta$ in the data with that in the prior.  Then combine the prior, scale, and likelihood/loss function to get a posterior for $\eta$ and, finally, get the corresponding marginal posterior for $\theta$, denoted by $\Pi_{n,\omega}$.  In addition to Bayes and Gibbs posteriors, this construction  includes variational Bayes posteriors, as discussed in the Supplementary Material, empirical Bayes, and others based on data-dependent priors \citep[e.g.,][]{martin.walker, hannig.2016}.  Our one technical requirement is that $\Pi_{n,\omega}$ be consistent in the sense that, under $P$, it concentrates around $\theta(P)$ asymptotically for each fixed $\omega$.  Consistency of $\Pi_{n,\omega}$ is not automatic, especially in cases of under- or misspecified models \citep[e.g.,][]{grunwald.ommen.scaling}, but it is necessary if credible regions derived from it are to provide meaningful uncertainty quantification.

For concreteness, consider inference on the median $\theta$ of a distribution $P$.  The median can be defined as the minimizer of the risk $R(\theta) = P \ell_\theta$, with loss function $\ell_\theta(z) = |z - \theta|$.  This loss forms a connection between $Z^n$ and $\theta$, and a Gibbs posterior is defined as 
\begin{equation}
\label{eq:Gibbs}
\Pi_{n,\omega}(d\theta) \propto e^{-\omega n R_n(\theta)} \, \Pi(d\theta), 
\end{equation}
where $R_n(\theta) = \PP_n \ell_\theta$ is the empirical version of the risk, $\omega > 0$ is a scale parameter, and $\Pi$ is a prior for $\theta$; other examples of converting risk into a Gibbs posterior are given in Sections~\ref{S:qreg} and \ref{S:application}.  As an alternative to the use of a loss function to connect data with the parameter of interest, one might specify a statistical model, e.g., $P$ is a gamma distribution with parameter $(\alpha, \beta)$.   This determines a likelihood function $L_n(\eta)$ which, combined with a prior $\Pi$ for $\eta$, yields a posterior for $\theta$ given by 
\[ \Pi_{n,\omega}(A) \propto \int_{\{\eta: F_\eta^{-1}(1/2) \in A\}} L_n(\eta)^\omega \, \Pi(d\eta), \]
where $F_\eta$ denotes the corresponding gamma distribution function.  The choice between these two approaches, or variations thereof, depends on the willingness of the data analyst to specify a full model, and on the objectives; the Gibbsian approach provides inference on the median but nothing else, with minimal modeling assumptions, whereas the Bayesian approach provides inference on virtually any feature, but with higher modeling and computational costs.  In either case, the choice of $\omega$ is important.  

Our proposed choice of scale is based on calibrating the posterior credible regions to be used for uncertainty quantification.  Fix a level $\alpha \in (0,1)$ and, for concreteness, consider the highest posterior density credible regions defined as 
\begin{equation}
\label{eq:credible}
C_{\omega, \alpha}(Z^n) = \{\theta: \pi_{n,\omega}(\theta) \geq c_\alpha\},
\end{equation}
where $\pi_{n,\omega}$ is the density function corresponding to the posterior $\Pi_{n,\omega}$, and $c_\alpha$ is chosen so that the $\Pi_{n,\omega}$-probability assigned to $C_{\omega, \alpha}(Z^n)$ is equal to $1-\alpha$.  Given that $\Pi_{n,\omega}$ is consistent, the scale parameter $\omega$ controls the spread of the posterior and, thus, the size of these credible regions.  Our proposal is to choose $\omega$ so that the credible regions for $\theta$ are of the appropriate size to be calibrated, i.e., so that the coverage probability 
%, $P\{C_{\omega, \alpha}(Z^n) \ni \theta(P)\}$, 
is approximately equal to $1-\alpha$.

\section{Posterior calibration}
\label{S:algorithm}

\subsection{Algorithm}

Our goal is to select $\omega$ such that the corresponding posterior credible region are calibrated in the sense that the credibility level agrees with the coverage probability, at least approximately.  To this end, for our desired significance level $\alpha \in (0,1)$, and our preferred credible region $C_{\omega, \alpha}(Z^n)$ for $\theta$ as in \eqref{eq:credible}, define the coverage probability function 
\[ c_\alpha(\omega; P) = P\{C_{\omega, \alpha}(Z^n) \ni \theta(P)\}, \]
i.e., the $P$-probability that the credible region $C_{\omega, \alpha}(Z^n)$ contains the target $\theta(P)$.  Then calibration requires that $\omega$ be such that 
\begin{equation}
\label{eq:calibrated}
c_\alpha(\omega; P) = 1-\alpha,
\end{equation}
i.e., that the $100(1-\alpha)$\% posterior credible region is also a $100(1-\alpha)$\% confidence region.  In practice we cannot solve this equation because we do not know $P$.  The approach described below is designed to get around this roadblock.  

To build up our intuition, start by assuming that $P$ is known.  Even in this case, numerical methods are generally required to solve \eqref{eq:calibrated}.  One can use stochastic approximation \citep[e.g.,][]{robbinsmonro} with iterations
\begin{equation}
\label{eq:sa}
\omega^{(t+1)} = \omega^{(t)} + \kappa_t\{\hat c_\alpha(\omega^{(t)} \mid P) - (1-\alpha)\}, \hspace{1cm} t\geq 0,
\end{equation}
where $\hat c_\alpha(\omega \mid P)$ is a Monte Carlo approximation to the coverage probability, obtained by simulating new copies of the data $Z^n$ from $P$, and $(\kappa_t)$ is a non-stochastic sequence such that $\sum_t \kappa_t = \infty$ and $\sum_t \kappa_t^2 < \infty$; in our examples we use $\kappa_t = (t+1)^{-0.51}$.  If $c_\alpha(\omega; P)$ is continuous and monotone decreasing in $\omega$, as would be expected even for moderate $n$ under our consistency assumption, at least on an interval away from 0, then the main result in \citet{robbins1971} implies $\omega^{(t)} \to \omega^\star$ $P$-almost surely, as $t \to \infty$, where $\omega^\star$ is the solution to \eqref{eq:calibrated}.

When $P$ is unknown, the proposed approach changes in two ways.  First, since it is not possible to sample new copies of $Z^n$ from $P$, we replace simulation from $P$ with that from $\PP_n$, i.e., we sample with replacement from the observed data $Z^n$.  Second, since we also do not know $\theta(P)$, we cannot check if a given credible region $C_{\omega, \alpha}(Z^n)$ covers it, so we use $\theta(\PP_n)$ in place of $\theta(P)$.  This results in an empirical version of $c_\alpha(\omega; P)$, namely, $c_\alpha(\omega; \PP_n) = \PP_n\{C_{\omega, \alpha}(Z^n) \ni \theta(\PP_n)\}$, 
%\begin{equation}
%\label{eq:emp.cov}
%c_\alpha(\omega; \PP_n) = \PP_n\{C_{\omega, \alpha}(Z^n) \ni \theta(\PP_n)\}, 
%\end{equation}
and then our proposal is to find $\omega$ such that 
\begin{equation}
\label{eq:praccalibrated}
c_\alpha(\omega; \PP_n) = 1-\alpha.
\end{equation}  
In practice, we cannot evaluate $c_\alpha(\omega; \PP_n)$ either, but the bootstrap provides a Monte Carlo estimator, $\hat c_\alpha(\omega; \PP_n)$.  We can solve \eqref{eq:praccalibrated} using the stochastic approximation procedure described above for the known-$P$ case.  Collectively, the steps in Algorithm~\ref{algo:calibrate} make up our \emph{general posterior calibration} (GPC) algorithm.  R code for several examples is available at \url{https://github.com/nasyring/GPC}.  

\begin{algorithm*}[t]
\smallskip
Fix a convergence tolerance $\eps > 0$ and an initial guess $\omega^{(0)}$ of the calibration parameter.  Take $B$ bootstrap samples $\tilde Z_1^n,\ldots,\tilde Z_B^n$ of size $n$.  Set $t=0$ and do:
\begin{enumerate}
\item Construct credible regions $C_{\omega^{(t)}, \alpha}(\tilde Z_b^n)$ for each $b=1,\ldots,B$.
\vspace{-2mm}
\item Evaluate the empirical coverage probability $\hat c_\alpha(\omega^{(t)}; \PP_n)$. 
\vspace{-2mm}
\item If $\bigl|\hat c_\alpha(\omega^{(t)}; \PP_n) - (1-\alpha)\bigr| < \eps$, then stop and return $\omega^{(t)}$ as the output; otherwise, update $\omega^{(t)}$ to $\omega^{(t+1)}$ according to \eqref{eq:sa}, set $t \gets t+1$, and go back to Step~1.
\end{enumerate}
\caption{\bf--- General Posterior Calibration.}
\label{algo:calibrate}
\end{algorithm*}

In most applications, the credible regions $C_{\omega,\alpha}(Z^n)$ are unavailable in closed form, so posterior sampling will be needed.  In the basic version of the algorithm presented here, this posterior sampling must be repeated for each bootstrap sample and each GPC iteration, which may or may not be prohibitive in a particular application.  But there are modifications that can be made to speed up the algorithm's performance.  For example, the bootstrap computations of $\hat c_\alpha(\omega; \PP_n)$ can be done in parallel, and we have implemented such a routine for the problem in Section~\ref{S:qreg}.  There, with $n=100$ data points, $B=200$ bootstraps, and $M=2000$ posterior samples, our parallelized implementation met the calibration criterion in less than 5 seconds on an ordinary desktop computer.  In addition, the frequency of posterior sampling can be reduced by adopting an importance sampling strategy when the change $\omega^{(t)} \to \omega^{(t+1)}$ is small.

\subsection{Theoretical support}

There are two theoretical questions of interest.  First, does a solution $\omega^\star$ to \eqref{eq:calibrated} exist?  Second, do the iterates from the proposed algorithm converge to $\omega^\star$?  The discussion here will focus primarily on the first question, but this analysis will also shed light on the second question and on some other existing methods for setting the scale $\omega$.  

We start with a simple case to build some intuition.  Let $Z^n$ be an independent sample of size $n$ from a normal distribution with mean $\theta$ and variance $\psi^2$; the goal is inference on the mean.  However, suppose that we fix the variance at $\sigma^2 \neq \psi^2$, i.e., a misspecified model.  If we construct a  posterior $\Pi_{n,\omega}$ using a scaling parameter $\omega$ and a flat prior, then the credible intervals are of the standard form: $\bar Z \pm z_\alpha (\omega^{-1} \sigma^2 n^{-1})^{1/2}$, where $z_\alpha$ is the upper $\alpha$ quantile of the standard normal distribution.  Clearly, to achieve the desired calibration, one must take $\omega = \sigma^2/\psi^2$.  Therefore, at least in this simple example, there is a solution to \eqref{eq:calibrated}, and note that it does not depend on $\alpha$.  Similar conclusions can be reached more generally and from different perspectives, as we discuss below.  

Next, suppose the $d$-dimensional $\theta=\theta(P)$ is defined via a risk function $R(\theta) = P\ell_\theta$, so that $\theta(P) = \arg\min R(\theta)$.  Then a Gibbs posterior $\Pi_{n,\omega}$ is defined as in \eqref{eq:Gibbs}, where $R_n(\theta) = \PP_n \ell_\theta$ is the empirical risk.  Under suitable regularity conditions \citep[e.g.,][]{chernozhukov.2003, muller2013}, the Gibbs posterior will be approximately normal, centered at $\theta(\PP_n) = \arg\min R_n(\theta)$, with asymptotic covariance matrix $\omega^{-1} \Sigma_n$, where $\Sigma_n = (n V_{\theta(\PP_n)})^{-1}$ and $V_\theta$ is the second derivative matrix of $R(\theta)$.  So, asymptotically, the $100(1-\alpha)$\% credible region $C_{\omega,\alpha}(Z^n)$ is
\[ \bigl[\theta: \omega\{\theta - \theta(\PP_n)\}^\top \Sigma_n^{-1} \{\theta-\theta(\PP_n)\} \leq q_\alpha \bigr], \]
where $q_\alpha$ is the upper $\alpha$ quantile of a chi-square distribution with $d$ degrees of freedom.  And for the above credible region, the coverage probability function is 
\[ c_\alpha(\omega; P) = G_n(q_\alpha / \omega), \]
where $G_n$ is the distribution function of $\{\theta(\PP_n) - \theta(P)\}^\top \Sigma_n^{-1} \{\theta(\PP_n) - \theta(P)\}$ under independent sampling of $Z^n$ or, equivalently, $\PP_n$ from $P$.  This asymptotic representation of the coverage probability as a smooth non-increasing function of $\omega$ reveals that a solution to \eqref{eq:calibrated} exists, at least for sufficiently large $n$.  

To see what the solution $\omega$ looks like, we push this argument further.  Under the same regularity conditions, the M-estimator $\theta(\PP_n)$ has an asymptotic representation 
\[ \theta(\PP_n) = \theta(P) + \Psi_n^{\text{c}} \, \xi + o_P(1), \]
where $\xi$ is a vector of independent standard normals and $\Psi_n^{\text{c}}$ is the Cholesky factor of the sandwich covariance matrix 
$\Psi_n = n^{-1} V_{\theta(P)}^{-1} \{ P \dot\ell_{\theta(P)} \dot\ell_{\theta(P)}^\top \} V_{\theta(P)}^{-1}$, 
with $\dot\ell_\theta$ the derivative of $\theta \mapsto \ell_\theta$.  This implies that
\[ \bigl| G_n(q_\alpha/\omega) - {\rm pr}(\xi^\top \Psi_n^{\text{c}\top} \, \Sigma_n^{-1} \Psi_n^{\text{c}} \, \xi \leq q_\alpha / \omega) \bigr| \to 0, \quad n \to \infty,  \]
pointwise and also uniformly in $\omega$, at least over a set bounded away from 0.  Therefore, if the generalized information equality \citep{chernozhukov.2003} holds, i.e., if $\Sigma_n = b\Psi_n$  
%$V_{\theta(P)} = P\dot\ell_{\theta(P)} \dot\ell_{\theta(P)}$, 
%then the asymptotic solution to \eqref{eq:calibrated} is $\omega=1$.  More generally, if $\Sigma_n = c \Psi_n$ 
%$V_{\theta(P)} = c P\dot\ell_{\theta(P)} \dot\ell_{\theta(P)}$ 
for a constant $c > 0$, then \eqref{eq:calibrated} holds asymptotically with $\omega=b$; compare this to the $\sigma^2/\psi^2$ result in the toy example.  More generally, with no information proportionality property, \eqref{eq:calibrated} will hold asymptotically for some $\omega$ greater than the smallest eigenvalue of $\{\Psi_n^{\text{c}\top} \, \Sigma_n^{-1} \Psi_n^{\text{c}}\}^{-1}$.  Therefore, under general conditions, a solution to \eqref{eq:calibrated} exists, at least asymptotically and, as in the toy example above, it does not depend on $\alpha$.  

The second question, about whether our proposed algorithm identifies a solution, is more difficult to answer.  Individually, the workhorses of the GPC algorithm, namely, stochastic approximation, bootstrap, and Monte Carlo, are sound computational tools, but a simultaneous analysis of the three working in tandem is formidable.  However, if we take the posterior computations and stochastic approximation to be exact, focusing only on the bootstrap component, then the above analysis provides some insight.  That the coverage probability can be characterized, asymptotically, via a distribution function means that solving for $\omega$ is equivalent to finding a quantile of the distribution of an approximate pivot.  And it is well known \citep[e.g.,][p.~39--41]{davison.hinkley.1997} that the bootstrap approximation is valid for such tasks under very mild conditions.  Therefore, at least asymptotically, theory suggests that our algorithm produces calibrated credible regions.  Moreover, the numerical examples presented here and in the Supplementary Material suggest that GPC achieves calibration, exactly or conservatively, even in finite samples.  But a scaling approach like ours cannot correct a misspecified shape, and the same goes for \citet{holmes.walker.2016} and \citet{grunwald.ommen.scaling}, so our calibrated posterior will generally be too wide in some direction; see Section~\ref{SS:gp}.

Finally, as a reviewer pointed out, there are close connections between those $\omega$ values that achieve calibration and those in \citet{grunwald2012} and \citet{grunwald.mehta.rates} achieve fast rates of convergence.  This connection has several important implications.  In particular, it suggests that there may be a unified framework for generalized posteriors where, at least asymptotically, a single choice of scaling achieves all the desired inferential goals.  But calibration is more delicate than a fast convergence rate, and this difference manifests here in that the calibration theory requires setting $\omega$ exactly at a threshold analogous to that given in \citet{grunwald.mehta.rates}, whereas the SafeBayes algorithm in \citet{grunwald.ommen.scaling} plays it safe by setting the scale factor slightly less than that threshold.  In the finite-dimensional cases being considered here, there is no serious risk in setting $\omega$ at the threshold to achieve both calibration and optimal convergence rates.  However, in more complex infinite-dimensional cases, one should expect to risk sub-optimal rates in exchange for valid uncertainty quantification \citep[e.g.,][]{martin.horseshoe.discuss}.  These interesting observations deserve further investigation.

\ifthenelse{1=1}{}{
Here we give a heuristic argument to support our claims that the proposed algorithm is providing at least approximately calibrated posterior credible regions.  Let the $d$-dimensional $\theta=\theta(P)$ be defined via a risk function $R(\theta) = P\ell_\theta$, so that $\theta(P) = \arg\min R(\theta)$.  In this case, a Gibbs posterior $\Pi_{n,\omega}$ is defined as in \eqref{eq:Gibbs}, where $R_n(\theta) = \PP_n \ell_\theta$ is the empirical risk.  Under suitable regularity conditions \citep[e.g.,][]{chernozhukov.2003}, the Gibbs posterior will be approximately normal, centered at $\theta(\PP_n) = \arg\min R_n(\theta)$, with asymptotic covariance matrix $\omega^{-1} \Sigma_n$, where $\Sigma_n = (n V_{\theta(\PP_n)})^{-1}$ and $V_\theta$ is the second derivative matrix of $R(\theta)$.  So, asymptotically, the $100(1-\alpha)$\% credible region $C_{\omega,\alpha}(Z^n)$ will be
\[ \bigl\{\theta: \omega\{\theta - \theta(\PP_n)\}^\top \Sigma_n^{-1} \{\theta-\theta(\PP_n)\} \leq q_\alpha \bigr\}, \]
where $q_\alpha$ is the upper $\alpha$ quantile of a chi-square distribution with $d$ degrees of freedom.  Therefore, for large $n$, the coverage probability function satisfies 
\[ c_\alpha(\omega; P) \approx G_n(q_\alpha / \omega), \]
where $G_n$ is the distribution function of $\{\theta(\PP_n) - \theta(P)\}^\top \Sigma_n^{-1} \{\theta(\PP_n) - \theta(P)\}$ under independent sampling of $Z^n$ or, equivalently, $\PP_n$ from $P$.  This approximation reveals that solving for $\omega$ is equivalent to finding a quantile, and has two important implications.  First, it shows that a solution to \eqref{eq:calibrated} exists, at least for sufficiently large $n$.  Second, it indicates that the bootstrap approximation to be employed in practical applications, which is known to be valid for quantile estimation in regular problems, is valid in this context as well.  To see what the solution $\omega$ looks like, we push this argument further.  Under the same regularity conditions, the M-estimator $\theta(\PP_n)$ has an asymptotic representation 
\[ \theta(\PP_n) = \theta(P) + \Psi_n^{\text{ch}} \, \xi + o_P(1), \]
where $\xi$ is a vector of independent standard normals and $\Psi_n^{\text{ch}}$ is the lower-triangular Cholesky factor of the asymptotic covariance matrix $\Psi_n$, given by  
\[ \Psi_n = n^{-1} V_{\theta(P)}^{-1} \{ P \dot\ell_{\theta(P)} \dot\ell_{\theta(P)}^\top \} V_{\theta(P)}^{-1}, \]
with $\dot\ell_\theta$ the derivative of $\theta \mapsto \ell_\theta$.  This implies that
\[ \lim_{n \to \infty} \bigl| G_n(q_\alpha/\omega) - {\rm pr}(\xi^\top \Psi_n^{\text{ch}} \, \Sigma_n^{-1} \Psi_n^{\text{ch}} \, \xi \leq q_\alpha / \omega) \bigr| = 0, \]
pointwise {\color{red} and uniformly} in $\omega$.  Therefore, if the generalized information equality \citep{chernozhukov.2003} holds, i.e., if $V_{\theta(P)} = P\dot\ell_{\theta(P)} \dot\ell_{\theta(P)}$, then the asymptotic solution to \eqref{eq:calibrated} is $\omega=1$.  More generally, if $V_{\theta(P)} = c P\dot\ell_{\theta(P)} \dot\ell_{\theta(P)}$ for a constant $c > 0$, the \eqref{eq:calibrated} holds asymptotically with $\omega=c$.  Even more generally, with no information proportionality property,  \eqref{eq:calibrated} will hold asymptotically for some $\omega$ greater than the smallest eigenvalue of $\{\Psi_n^{\text{ch}} \, \Sigma_n^{-1} \Psi_n^{\text{ch}}\}^{-1}$.  Therefore, at least asymptotically, under regularity conditions, we expect our algorithm to produce calibrated credible regions.  However, a scaling approach like ours, including those in \citet{holmes.walker.2016} and \citet{grunwald.ommen.scaling}, cannot correct a misspecified shape, so our calibrated posterior will generally be too wide in some direction; see Section~\ref{SS:gp}.
}

\subsection{Illustration}
\label{SS:gp} 

Example~1 in \citet{ribatet.cooley.davison.2012} presents a case where $\{Y(x): x \in \RR\}$ is a Gaussian process with mean $\mu \in \RR$ and covariance function $K(x,x') = \tau e^{-\kappa |x-x'|}$, where $\kappa > 0$ is known but $\theta=(\mu,\tau)$ is unknown.  They present two Bayesian analyses, one based on the full likelihood and one based on a pairwise composite likelihood, the latter motivated by simpler computations.  But since the composite likelihood is misspecified in that it treats all pairs of observations as independent, the uncertainty measures it produces are overly optimistic, i.e., the posterior is too concentrated.  

\begin{figure}[t]
\begin{center}
\subfigure[Marginal density for $\mu$]{\scalebox{0.4}{\includegraphics{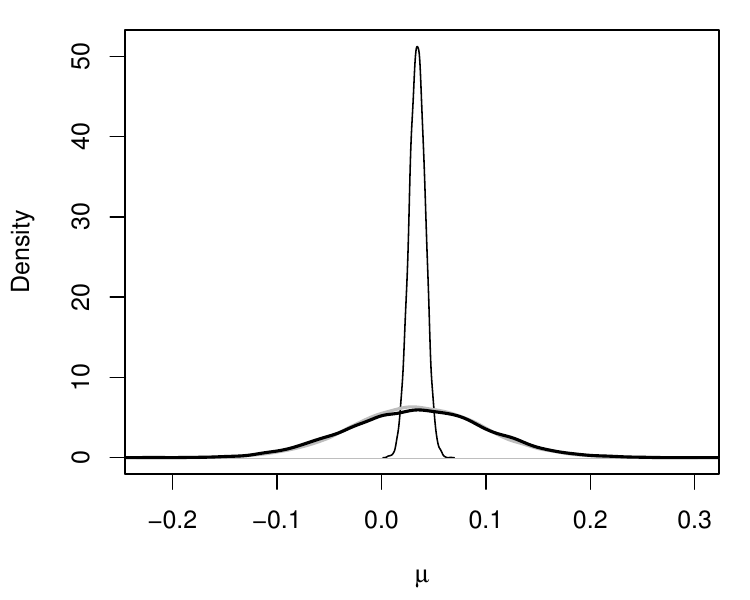}}}
\subfigure[Marginal density for $\tau$]{\scalebox{0.4}{\includegraphics{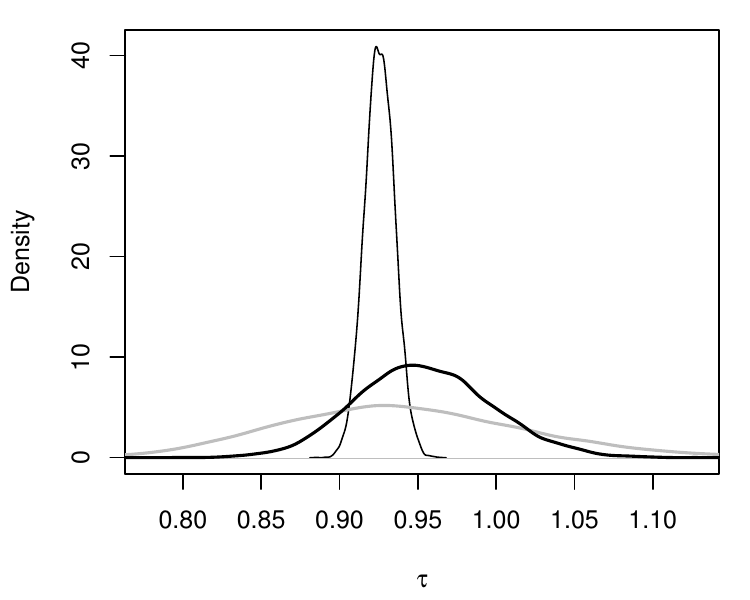}}}
\subfigure[Samples of $(\mu,\tau)$]{\scalebox{0.4}{\includegraphics{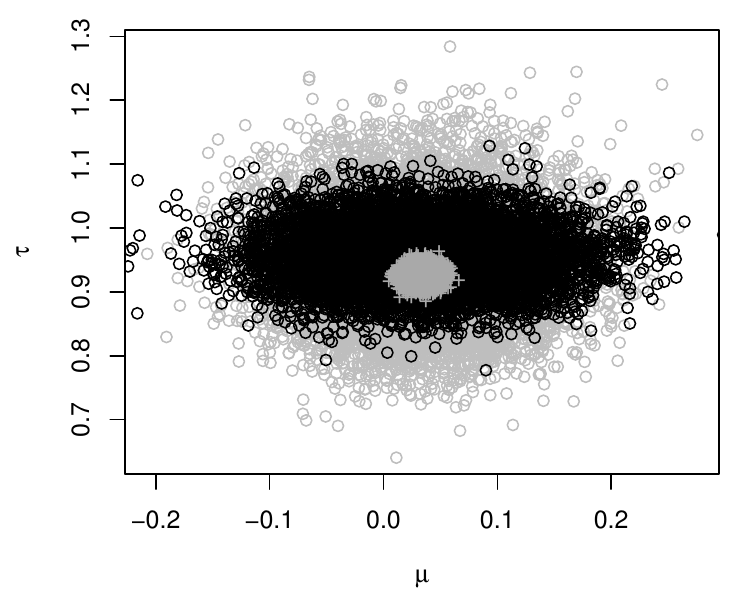}}}
\end{center}
\caption{Results for the example in Section~\ref{SS:gp}.  Panels~(a) and (b) show the Bayesian (thick black line), composite (thin black line), and scaled composite (gray line) posterior densities for $\mu$ and $\tau$, respectively.  In Panel~(c), tight gray cluster in the foreground is the pairwise composite posterior, black is the Bayesian posterior, and wider gray cluster in the background is the scaled composite posterior, with $\omega=0.015$.}
\label{fig:gp}
\end{figure}

To illustrate our scaling method, we reproduce the simulation summarized in Figure~1 of \citet{ribatet.cooley.davison.2012}.  Our Figure~\ref{fig:gp} summarizes the full posterior, the composite likelihood-based posterior, and our calibrated version of the latter, where the scaling parameter return by our posterior calibration algorithm is $\omega=0.015$.  The two marginal posterior density plots in Panels~(a) and (b) match what is shown in \citet{ribatet.cooley.davison.2012}.  Our calibrated posterior exactly matches the full posterior for $\mu$, but is a bit wider than that for $\tau$.  Panel~(c) summarizes the joint posteriors and helps explain what is going on.  The composite posterior is too tightly concentrated, but also is more circular than the full posterior.  Our scaling approach stretches the composite posterior's roughly circular contours till they achieve calibration, and these will necessarily exceed the narrowest part of the full posterior's elliptical contours, as in Panel~(c).

%\section{Numerical comparisons}
%\label{S:examples}

%\subsection{Quantile regression}
%\label{SS:qreg}

\section{Quantile regression example}
\label{S:qreg}

In quantile regression, for fixed $\tau \in (0,1)$, we are interested in the $\tau^{\text{th}}$ quantile of the response $Y\in \RR$, given the covariates $X \in \RR^{p+1}$, expressed as
\begin{equation}
\label{eq:qrmodel}
Q_\tau(Y \mid X) = X^\top\theta,
\end{equation}
where dimension $p+1$ represents an intercept and $p$ covariates.  In \eqref{eq:qrmodel}, the vector $\theta$ depends on $\tau$ but we will omit this dependence for simplicity.  Inference on the quantile regression coefficient $\theta$ may be carried out using asymptotic approximations \citep[][Theorem~4.1]{koenker.2005} or the bootstrap \citep{horowitz.1998}.  A Bayesian approach would also be attractive, but no distributional form is given in \eqref{eq:qrmodel} so a likelihood requires further specification.  An option considered by several authors \citep[e.g.,][]{yu.moyeed.2001, sriram.rvr.ghosh.2013, sriram.2015} is to use a misspecified asymmetric Laplace likelihood.  This corresponds to a Gibbs posterior \eqref{eq:Gibbs} using the empirical risk 
\begin{equation}
\label{eq:check}
R_n(\theta) = \frac1n \sum_{i=1}^n |(Y_i - X_i^\top\theta)(\tau - 1_{Y_i < X_i^\top\theta})|,
\end{equation}
where $1_A$ denotes the indicator function for $A$.  Here we consider $\tau=0.5$.  
 
%It follows from \citet{kleijn.2012} that the Gibbs posterior based on \eqref{eq:check} satisfies a Bernstein--von Mises theorem.  Despite the desirable convergence result, the variance mismatch discussed in Section~\ref{S:problem} causes the credible regions to be too large and over-cover.  On the other hand, the GPC  algorithm calibrates the intervals exactly, for all $n$, without loss of efficiency in terms of interval lengths.  

To demonstrate the performance of our proposed scaling algorithm, we revisit a simulation example presented in \citet{yang.he.2012}.  The model they consider is
\[Y_i = \theta_0 + \theta_1 \, X_i + e_i, \quad i=1,\ldots,n, \]
where $\theta_0 = 2$, $\theta_1 = 1$, $e_i \iid \nm(0,4)$, and $X_i + 2 \iid \chisq(2)$.  \citet{yang.he.2012} showed numerically that their Bayesian empirical likelihood method produced credible intervals with approximate coverage near the nominal 95\% level.  They also show their method produces credible intervals with shorter average lengths than a Gibbs posterior with $\omega$ equal to the average absolute residuals calculated using the usual quantile regression parameter estimates.  The results for these methods are presented in Table~\ref{table:qreg}, along with the results from the posterior intervals scaled by our algorithm.  

There are two key observations to be made.  First, our method calibrates the credible intervals to have exact 95\% coverage across the range of $n$, while the other methods tend to over-cover.  Second, our credible intervals tend to be shorter than those of the other methods, especially for $n=100$.  All three methods have a $n^{-1/2}$ convergence rate so, for large $n$, we cannot expect to see substantial differences between the various methods.  Therefore, the small-$n$ case should be the most important and, at least in this case, the credible intervals calibrated using our algorithm are the best.         

\begin{table}%[ht]
	\centering
\resizebox{\textwidth}{!}{\begin{tabular}{lcccccccccccc}
\hline
&  & \multicolumn{5}{c}{Coverage Probability $\times 100$} & & \multicolumn{5}{c}{Average Length $\times 100$} \\
\cline{3-7}
\cline{9-13}
$n$ & & BEL.s & BDL & Normal & $\omega \equiv 0.8$ & GPC & & BEL.s & BDL & Normal & $\omega \equiv 0.8$ & GPC \\
\hline
$100$ & $\theta_0$ & $97$& $98$ & 95 & 96 & 95 & & $106$ & $111$ & 100 & 100 & 91 \\
    & $\theta_1$ & $98$ & $98$ & 98 & 98 & 95 & & $58$ & $58$ & 55 & 52 & 47 \\
$400$ & $\theta_0$ & $95$ & $98$ & 95 & 95 & 95 & & $50$ & $55$ &50 & 49 &  46 \\
    & $\theta_1$ & $97$  & $98$ & 97 & 96 & 95 & & $26$ & $28$ &25 & 25 &  23 \\
$1600$& $\theta_0$ & $96$  &$97$ & 96 & 95 & 95 & & $25$ & $28$ & 25 & 24 & 23 \\
    & $\theta_1$ & $96$  & $98$ & 96 & 96 & 95 & & $13$ & $14$ & 12 & 12 & 11 \\
\hline
\end{tabular}}
\caption{Comparison of $95\%$ posterior credible intervals of the median regression parameters from five methods: BEL.s and BDL are two methods presented in \citet{yang.he.2012}; Normal is the confidence interval computed using the asymptotic normality of the M-estimator; $\omega \equiv 0.8$ is the scaled posterior with fixed $\omega$; and our GPC. Coverage probability and average interval lengths are based on $5000$ simulated data sets.  
}
\label{table:qreg}
\end{table}

Finally, considering that in smooth models we expect $\omega$ to account for the difference in asymptotic variance between the posterior and the M-estimator, it is reasonable to ask if we need a calibration algorithm at all, i.e., can we get by with a fixed value of $\omega$ based on these asymptotic variances?  A comparison of the asymptotic variance of the posterior with that of the M-estimator shows that $\Sigma_n \approx 0.80 \Psi_n$; therefore, we can take $\omega \equiv 0.80$ in an attempt to calibrate posterior credible intervals with a fixed scaling.  Table~\ref{table:qreg} shows that our algorithm is still better than using a fixed scale based on asymptotic normality, especially at smaller sample sizes where the normal approximation is less justifiable.

%\subsection{Linear regression}
%\label{S:lin_reg}

%\subsection{Finite normal mixtures}
%\label{S:varinf}

\section{Application}
\label{S:application}

\citet{polson.scott.2011} propose to convert the objective function of the  support vector machine into a sort of log-likelihood function for the purpose of carrying out a Bayesian analysis.  Let $y$ be a binary $n$-vector, with $y_i \in \{-1,+1\}$ for $i=1,\ldots,n$, and $X$ a $n \times p$ matrix of predictors, including a column of ones for an intercept.  Then the support vector machine seeks to find $\theta=(\theta_0,\ldots,\theta_{p-1}) \in \RR^p$ to minimize 
\[ D_n(\theta) = n R_n(\theta) + \nu^{-1} \sum_{j=0}^{p-1} \bigl| \theta_j / \sigma_j \bigr|, \]
where $\nu > 0$ is a tuning parameter, $\sigma_j$ is the standard deviation of the $j$th column of $X$, with $\sigma_0 \equiv 1$, and $R_n(\theta) = n^{-1}\sum_{i=1}^n 2 \max\{0, 1 - y_i x_i^\top \theta\}$, with $x_i^\top$ the $i$th row of $X$.  Then \citet{polson.scott.2011} propose a pseudo-posterior distribution $\Pi_n$ with density function proportional to $\exp\{-D_n(\theta)\}$, which amounts to combining a pseudo-likelihood $L_n(\theta) = \exp\{-n R_n(\theta)\}$ with independent Laplace-type prior, $\Pi$, for $\theta$.  Their motivation is that, while the support vector machine only provides point estimates, this Bayesian formulation also offers uncertainty quantification.  However, as this support vector machine-driven posterior has no connection to a model that describes the variability in the data, it is not clear what the uncertainty measures derived from this posterior represent; certainly there is no reason to expect that posterior credible regions derived from it will be calibrated in the sense considered here.  To overcome this, we can introduce the scale parameter $\omega$, get a posterior $\Pi_{n,\omega}$ for $\theta$, and then $\omega$ can be chosen according to the calibration algorithm proposed above, thereby calibrating the corresponding credible region.  

\begin{figure}[t]
\begin{center}
\subfigure[Tobacco use, $\theta_1$]{\scalebox{0.4}{\includegraphics{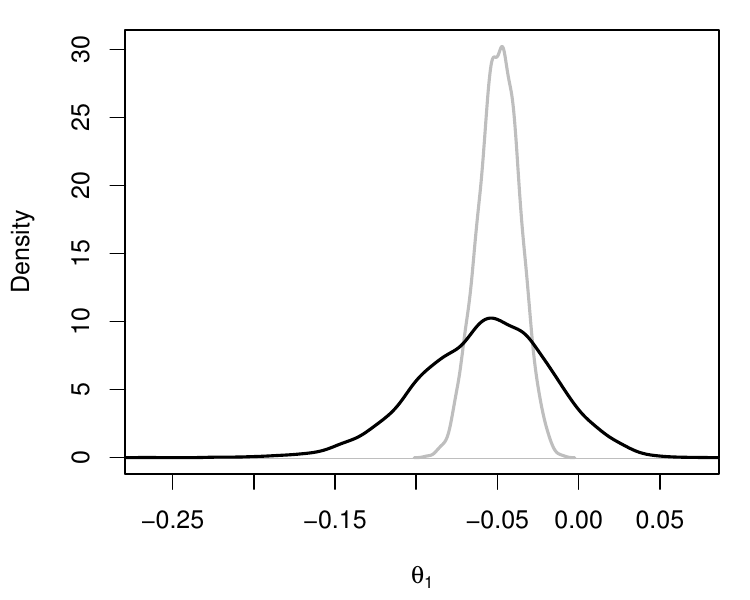}}}
\subfigure[Cholesterol, $\theta_2$]{\scalebox{0.4}{\includegraphics{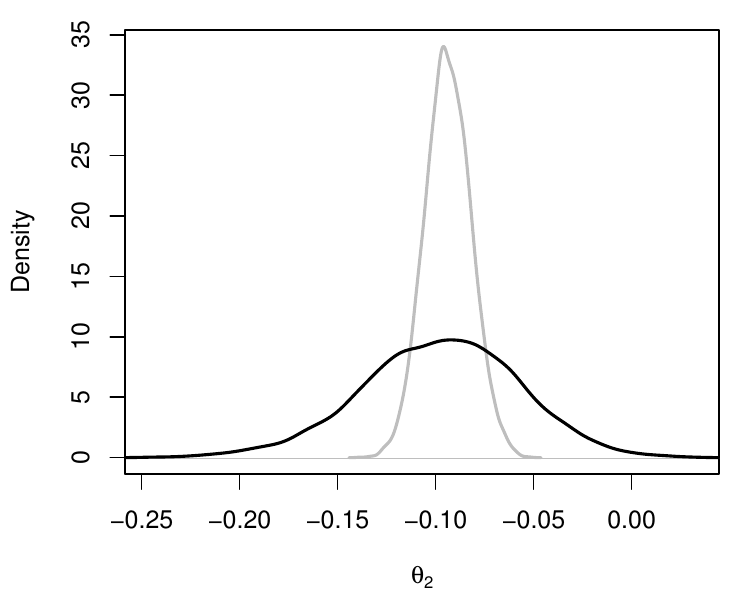}}}
\subfigure[Family history, $\theta_3$]{\scalebox{0.4}{\includegraphics{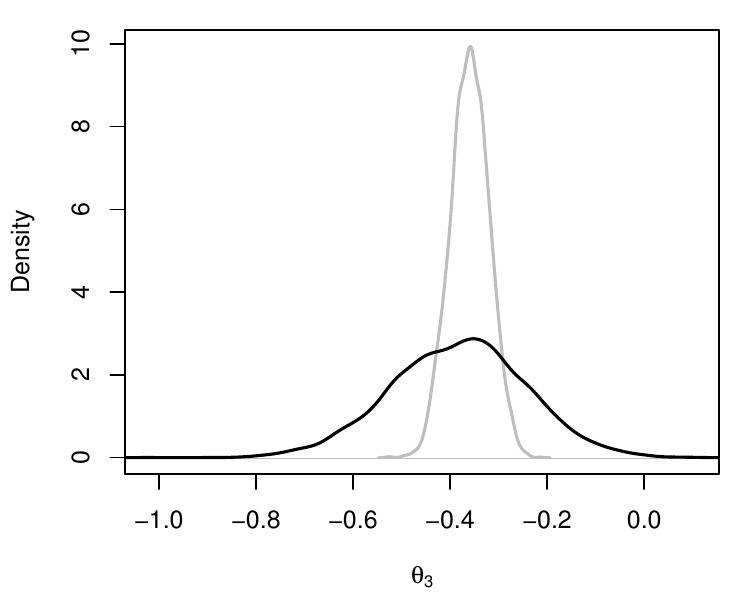}}}
%\subfigure[Age]{\scalebox{0.5}{\includegraphics{heart_beta4}}}
\end{center}
\caption{Marginal posterior densities in the support vector machine-based model; \citet{polson.scott.2011}, gray, and our calibrated version, black. 
}
\label{fig:heart}
\end{figure}

To illustrate this, consider the South African heart disease data presented analyzed in Section~4.4.2 of \citet{hastie.tibshirani.friedman.2009}.  The binary response is the presence/absence of myocardial infarction.  We focus here on a subset of predictors, in Table~4.3 of \citet{hastie.tibshirani.friedman.2009}, determined to have a non-negligible association with the response, namely, tobacco use, cholesterol level, family history of heart disease, and age.  The Gibbs sampler in \citet{polson.scott.2011} can be easily modified to incorporate our scaling $\omega$, and our calibration algorithm yields $\omega \approx 0.09$, for a target 95\% coverage; here we fix $\nu=10$.  Figure~\ref{fig:heart} shows the estimated marginal density functions for each of $\theta_1$, $\theta_2$, and $\theta_3$ based on the Polson and Scott proposal, with $\omega=1$, and ours with $\omega=0.09$; the plot for $\theta_4$ looks similar, so is omitted for the sake of space.  These plots indicate that, compared to our suitably calibrated posterior, Polson and Scott's posterior is far too narrow, exaggerating the precision of their inferences.

\ifthenelse{1=1}{}{
\section{Concluding remarks}
\label{S:discuss}

The sensitivity of Bayesian credible sets to the posited probability model makes obtaining calibrated inference a challenging problem.  Our linear regression example demonstrates this sensitivity when we take the model for granted.  However, misspecification can happen in a variety of settings, and not always unintentionally.  In quantile regression, the model is determined by a risk function rather than a likelihood, making traditional Bayesian inference using the true likelihood elusive.  And, other times, computational considerations make composite likelihood or variational posteriors an attractive alternative to a fully Bayesian analysis.  Our posterior calibration algorithm may provide a satisfactory solution in all of these settings by correcting model misspecification in a certain sense, providing calibrated inferences, at least approximately.     

Although the focus in this paper is on models that are misspecified in some way, it may still be desirable to apply our algorithm even in correctly specified cases.  That is, exact probability matching requires a certain choice of prior distribution, but our algorithm allows the user to pick any prior they desire, even an informative one, and tweak the posterior in a suitable way to achieve posterior calibration.

%Finally, while it is clear that the GPC algorithm produces approximately calibrated credible sets, a detailed theoretical study is needed.  The techniques we have used---stochastic approximation, MCMC, and bootstrap---each are theoretically sound on their own, but very complicated when used in tandem.  Further work in this direction may help provide guidance, but the lack of completely rigorous theory does not take away from the encouraging examples shown throughout the paper.

Finally, there are interesting follow-up projects in a variety of directions.  In particular, important application areas include spatial extremes, variance component models, etc., and more sophisticated sampling methods should be investigated to reduce the computational cost of the basic implementation of the algorithm employed here. 

%Our algorithm is a promising first step towards calibrating posteriors, but further work is needed. First, the techniques employed in our implementation of our algorithm are perhaps too simple to be efficient in more complex problems. For example, more sophisticated versions of the resampling strategy may make the approximations more accurate or faster to compute. For instance, we use Metropolis--Hastings MCMC in our linear and quantile regression examples, but sampling from variational approximations to the posterior are a possible alternative \citep{alquier.2015}. Alternatively, MCMC could be used to sample the posterior for the initial value of $\omega$ and then sampling importance resampling could be used to resample from the posterior samples as $\omega$ is updated. It would also be interesting to explore if any of the steps in our algorithm could be done in parallel to speed up the computations. Second, the examples shown in this paper consist of problems involving only a few parameters, so efforts are needed to make the posterior sampling as well as calibrating more efficient. 
}

\section*{Acknowledgments}

The authors thank the Editor, Associate Editor, and anonymous reviewers for their helpful feedback on an earlier version of this paper.  This work is partially supported by the U.~S.~Army Research Offices, Award \#W911NF-15-1-0154.

\bibliographystyle{apalike}
\bibliography{mybib_c}

\bigskip

\hrule

%\bigskip

%\pagebreak

\section*{Supplementary material}

\subsection*{Linear regression example}

Consider the linear regression model for data $(X_i, Y_i) \in \RR^p \times \RR$,
\begin{equation}
\label{mis:linreg}
Y_i = \beta_0 + X_i^\top \beta + \sigma \, e_i, \quad i=1,\ldots,n,
\end{equation}
where $\beta \in \RR^p$ is the vector of slope coefficients, $\sigma > 0$ is an unknown scale parameter, and $e_1,\ldots,e_n$ are assumed to be independent and identically distributed normal random variables with mean $0$ and variance $1$.  Suppose, however, that the constant error variance assumption is violated, in particular, the variance of $e_i$ is $\|X_i\|$, $i=1,\ldots,n$.  Our choice of predictor-dependent variance is a less-stylized version of that in \citet{grunwald.ommen.scaling}.  The proposed model is, therefore, misspecified, but our goal is still to obtain calibrated inference on $\theta = (\beta_0, \beta)$.  

The Jeffreys prior with density $\pi(\eta) \propto (\sigma^2)^{-3/2}$ is a reasonable default choice \citep{ibrahim1991} for the full parameter $\eta = (\theta, \sigma^2)$.  Since this prior is probability-matching for the location-scale model \citep[e.g.,][]{datta.2004}, we may expect that the posterior credible intervals would be approximately calibrated for our linear regression.  However, for a misspecified model, calibration might fail; in fact, as shown in Table~\ref{tbl:linmod}, the credible intervals are too narrow and tend to undercover.  

To investigate the performance of our proposed posterior calibration method, we carry out a simulation study.  We simulated data sets of $n=50$ observations.  Each $X_i \in \RR^3$ is multivariate normal with zero mean and unit variance for each element, and correlation $0.5$ for $X_{i1}$ and $X_{i2}$ and zero otherwise.  To sample $Y_i$ we use $\beta_0 = 0$, $\beta = (1,2,-1)^\top$, and $\sigma = 1$.  Although the error variance contains $\|X_i\|$, the White test for constant variance does not detect the heteroscedasticity reliably.  Table~\ref{tbl:linmod} shows the estimated coverages and mean lengths of several posterior credible intervals for the components of $\theta$.  Besides those scaled by the general posterior calibration algorithm, we consider a misspecified Bayes approach that fixes $\omega \equiv 1$, and posteriors with scale $\omega$ chosen by the method in \cite{holmes.walker.2016} and the method in \citep[][Algorithm~1]{grunwald.ommen.scaling}.  Table~\ref{tbl:linmod} shows that for this example {SafeBayes} performs similarly to general posterior calibration, while the method in \cite{holmes.walker.2016} does not improve upon the misspecified Bayesian model in terms of calibration.

To perform general posterior calibration we begin by fitting the linear model to the data and generating $B = 200$ bootstrap resampled data sets, $(Y_b^\star, X_b^\star)$ for $b = 1, ..., B$, by selecting $n$ rows with replacement from $Y$ and $X$.  For each bootstrap sample, $(Y_b^\star, X_b^\star)$, we fit the linear model and produce bootstrap estimates $\beta_{0,b}^\star$ and $\beta_{b}^\star$.    We set $\omega = \omega_0$ and use a Gibbs sampler to generate $M = 2000$ samples from the posterior distribution of $(\theta, \sigma^2)$, given $\omega$ and $(Y_b^\star, X_b^\star)$ for each of the $B$ bootstrap resampled data sets.  From these $B$ sets of $M$ posterior samples, we compute $B$ credible intervals for $\beta_0$ and each element of $\beta$.  Simple equi-tailed credible intervals can be used, but we found that highest posterior density credible intervals are more accurate in practice.  The average of the bootstrap estimates, $\tilde{\beta} = B^{-1}\sum_{b=1}^B \hat{\beta}^\star_b$, is used in place of the unknown $\beta$ to determine coverage proportions for each posterior credible interval.  Then 
\[ \hat c_\alpha(\omega; \PP_n) = \frac{1}{B} \sum_{b=1}^B 1\{C_{\omega, \alpha}(Y_b^\star, X_b^\star) \ni \tilde{\beta}\},\] 
where $C_{\omega, \alpha}(Y_b^\star, X_b^\star)$ is the highest posterior density credible set for $\beta$.  Then, the stochastic approximation step is used to update $\omega$ according $(4)$.  The algorithm converges when $\hat c_\alpha(\omega; \PP_n) - (1-\alpha) < \eps$, and we take $\eps = B^{-1}$.  

The implementation of general posterior calibration can be modified to increase accuracy, decrease runtime, or perhaps both.  In this linear model example, general posterior calibration runs faster for a Gibbsian posterior based on the empirical risk function $R_n(\beta_0, \beta) = n^{-1} \sum_{i=1}^n(Y_i - \beta_0 - X_i^\top\beta)^2$ than for the misspecified Bayes posterior due to the lack of the nuisance variance parameter $\sigma^2$ in the former.  

In general, the posterior calibration algorithm can be accelerated by decreasing $M$, $B$, or both, but reducing either number will also reduce the quality of the empirical coverage proportions $\hat c_\alpha(\omega; \PP_n)$.  In our implementation of general posterior calibration, the posterior is sampled every time $\omega$ is updated.  However, it may be faster to sample the posterior $M$ times for $\omega_0$ and subsequently use importance sampling to update the posterior samples each time $\omega$ is updated.  

The above simulation was repeated with $\kappa_t = ct^{-0.51}$ for $c = 1, 2,$ and $5$.  While there were no appreciable differences in credible interval coverage proportions or lengths for the different choices of $\kappa_t$, the algorithm ran in about 40, 30, and 50 seconds with $c=1$, $c = 2$, and $c = 5$, respectively, suggesting that an optimal choice of $\kappa_t$ could improve general posterior calibration runtime.  These run-times could also be reduced via parallelization, as we did for the example in Section~4 of the main text.  

We used our general posterior calibration algorithm to select $\omega$ to calibrate $95\%$ credible regions, simultaneously, for all three elements of $\beta$.  However, we also calibrate $100(1-\alpha)\%$ credible intervals for any confidence level $\alpha$ with the same $\omega$.  For instance, $90\%$ and $80\%$ credible intervals were also calibrated to a similar degree of accuracy compared with $95\%$ credible intervals.  This confirms the theoretical claims made in Section~3.2 of the main text.  

%Abusing notation slightly and appending a column of ones to the first column of the design matrix $X$ to accommodate the intercept $\beta_0$, it is easy to show that $\partial^2 \ell / \partial \beta_j \partial \beta_k = \sum_{i=1}^n X_{ij} X_{ik}/\sigma^2$ for $j,k \in \{0,1,2,3\}$ where $\ell(\sigma^2, \beta | (Y, X))$ denotes the likelihood for the misspecified model.  The mixed partial derivatives $\partial^2 \ell / \partial \sigma^2 \partial \beta_j  \approx 0$.  Similarly, the off-diagonal entries of $P\dot{\ell}\dot{\ell}^\top$ in the first row and column are zero, while the entries corresponding to derivatives with respect to $\beta_j$ and $\beta_k$ equal $\sum_{i=1}^n ||X_i||X_{ij} X_{ik}/\sigma^4$.  Therefore, the relevant comparison is between the entries $\sum_{i=1}^n X_{ij} X_{ik}/\sigma^2$ of $V_{\sigma^2,\beta}$ and the entries $\sum_{i=1}^n ||X_i||X_{ij} X_{ik}/\sigma^4$ of $P\dot\ell \dot\ell^\top$.  Since there is no constant $c$ for which $V_{\sigma^2,\beta} = cP\dot\ell \dot\ell^\top$, by the argument given in Section~3.2, any $\omega$ larger than the smallest eigenvalue of $\{\Psi_n^{\text{chol}} \, \Sigma_n^{-1} \Psi_n^{\text{chol}}\}^{-1}$ will calibrate credible intervals asymptotically, for all values of $\alpha \in (0,1)$, simultaneously.        

It is interesting to compare $\omega$ values generated by different algorithms.  The mean $\omega$ values are 0.80 and 0.77 for general posterior calibration and Gr\"unwald--van Ommen.  General posterior calibration is less variable than Gr\"unwald--van Ommen with standard deviation 0.05 for the former and 0.15 for the latter.  The method in Holmes--Walker generates $\omega$ values that average $1.29$, significantly different than those from general posterior calibration and Gr\"unwald--van Ommen.     

\begin{table}[t]
\centering
\begin{tabular}{lllcccc}
\hline
                                    %&          &  & \multicolumn{4}{c}{Multiple Regression}                              \\ \cline{4-7} 
                                    &          &  & $\beta_0$  & $\beta_1$  & $\beta_2$  & $\beta_3$ \\ \hline
\multirow{2}{*}{Misspecified Bayes} & coverage &  & 94       & 89       & 88       & 87                          \\
                                    & length   &  & 99& 116 & 116 & 101                    \\ 
\multirow{2}{*}{General posterior calibration}             & coverage &  & 98       & 94       & 94       & 93                          \\
                                    & length   &  & 117 & 136& 136 & 118                    \\
\multirow{2}{*}{SafeBayes}          & coverage &  & 96       & 93       & 94       & 92                          \\
                                    & length   &  & 119 & 140 & 139& 121                    \\ 
\multirow{2}{*}{Holmes and Walker}  & coverage &  & 91       & 84       & 80       & 82                          \\
                                    & length   &  & 87 & 101& 101 & 87                    \\ \hline
\end{tabular}
\caption{Empirical coverage probability of $95\%$ credible intervals and average interval lengths calculated using 5000 simulations from the linear regression model.  All values are multiplied by $100$.}
\label{tbl:linmod}
\end{table}

\subsection*{Finite mixture model example}

Variational inference offers a competing method to Markov chain Monte Carlo for approximating posterior distributions.  This approach specifies a family of distributions, often a normal family, as candidate posteriors and then chooses the parameters of that family to minimize the Kullback--Leibler divergence from the true posterior.  The variational posterior is simple by construction and, if carefully chosen, will be consistent \citep[e.g.,][]{wang.titter.2005}, but as noted in \citet{blei.VI}, misspecification causes the variational posterior variance to be too small.  

As an example, we consider the normal mixture model presented in \citet{blei.VI}, i.e., $Y_1,\ldots,Y_n$ are independent and identically distributed from  
\begin{equation}
\label{eq:normal.mix}
\sum_{k=1}^K \pi_k \nm(\mu_k, \sigma_k^2). 
\end{equation}
The full parameter $\eta$ consists of the mixture weights $(\pi_1,\ldots,\pi_K)$, means $(\mu_1,\ldots,\mu_K)$, and variances $(\sigma_1^2, \ldots, \sigma_K^2)$, but we will consider inference only on the means.  We can construct a variational posterior for $\eta$ following Algorithm~2 in \citet{blei.VI}, which approximates the posterior by a multivariate normal distribution.  The additional scale factor $\omega$ in our modified variational posterior $\Pi_{n,\omega}$ only adjusts the overall scale of this multivariate normal distribution.  Therefore, if $m_1,\ldots,m_K$ and $v_1,\ldots,v_K$ are the means and variances, respectively, of this variational posterior for the mixture means $\mu_1,\ldots,\mu_K$, then the corresponding $\omega$-scaled variational posterior $100(1-\alpha)$\% credible intervals are of the form 
\[ \mu_k \pm z_{\alpha / 2}^\star \, \omega \, v_k^{1/2}, \quad k=1,\ldots,K. \]
It is straightforward to apply our general posterior calibration algorithm to variational posteriors; the computational investment is in carrying out the optimization needed for the variational  approximation at each bootstrap step, but then the credible intervals are available in closed-form so no posterior sampling is needed.

We claim that the general posterior calibration algorithm will properly scale the variational posterior, calibrating the corresponding credible intervals, correcting the under-estimation of variance noted in \citet{blei.VI}.  To demonstrate this, we carry out a simple simulation study.  We take $K=2$, $\pi_1 = \pi_2 = 1/2$, $(\mu_1, \mu_2) = (-2, 2)$, and $\sigma_1 = \sigma_2 = 1$.  Table~\ref{tbl:VI} shows the empirical coverage probabilities and mean lengths of the 95\% credible intervals based on Algorithm~2 in \citet{blei.VI} and our general posterior calibration algorithm.  Apparently, our algorithm corrects the underestimated variance of the variational posterior, producing credible intervals that are slightly conservative.  

\begin{table}
\centering
\begin{tabular}{lllccc}
\hline
                     %&          &  & \multicolumn{3}{c}{Mixture Means} \\ \cline{4-6} 
                     &          &  & $\mu_1$      &      & $\mu_2$     \\ \hline
\multirow{2}{*}{General posterior calibration} & coverage &  &  96        &      &  96       \\
                     & length   &  &  67   &      &  67  \\
\multirow{2}{*}{Variational posterior}  & coverage &  &  92        &      &  92       \\
                     & length   &  &  55   &      &  55  \\ \hline 
\end{tabular}
\caption{Empirical coverage probabilities and average lengths of the 95\% credible intervals for $(\mu_1, \mu_2)$ based on our general posterior calibration algorithm and the variational posterior in \citet{blei.VI} over 5000 simulated data sets from the mixture model \eqref{eq:normal.mix}.  All results are multiplied by $100$.}
\label{tbl:VI}
\end{table}

\end{document}